\documentclass[structabstract]{aa}  

\usepackage[varg]{txfonts}
\usepackage{verbatim}
\usepackage{amssymb}
\usepackage{natbib}
\bibpunct{(}{)}{;}{a}{}{,}

\newcommand{\amm}{NH$_3$}
\newcommand{\wat}{H$_2$O}
\newcommand{\met}{CH$_3$OH}

\newcommand{\kms}{km~s$^{-1}$}

\newcommand{\cmc}{cm$^{-3}$}

\newcommand{\ls}{$L_{\odot}$}

\def\kmsy{km~s$^{-1}$~yr$^{-1}$}

\newcommand{\pas}{$\rlap{.}^{\prime\prime}$}
\newcommand{\degree}{$^{\circ}$}

\newcommand{\tkin}{T$_{\rm kin}$}

\begin{document}

\title{Hot Ammonia around Young O-type Stars}
\subtitle{II. JVLA imaging of highly-excited metastable \amm\,masers in W51-North}

\author{ C. Goddi \inst{1}
\and C. Henkel \inst{2,3}
\and Q. Zhang  \inst{4}
\and L. Zapata \inst{5}
\and T. L. Wilson \inst{6}
}
\institute{Joint Institute for VLBI in Europe, Postbox 2, 7990 AA Dwingeloo, The Netherlands  
\and
Max-Planck-Institut  f\"{u}r Radioastronomie, Auf dem H\"{u}gel 69, 53121 Bonn, Germany
\and
Astronomy Department, Abdulaziz University,
                   P.O. Box 80203, Jeddah 21589, Saudi Arabia
\and                     	
   Harvard-Smithsonian Center for Astrophysics, 60 Garden Street, Cambridge, MA 02138
\and
Centro de Radioastronom\'{i}a y Astrofis\'{i}ca, UNAM, Morelia, Michoac\'{a}n, M\'{e}xico, C.P. 58089
\and
US Naval Research Laboratory, Code 7213, Washington, DC 20375, USA
}

\abstract
{This paper is the second in a series of ammonia (NH$_3$) multilevel imaging studies in high-mass star forming regions. }
{  We want to identify the location of the maser emission from highly-excited levels of ammonia within the  W51~IRS2 high-mass star forming complex, that was previously discovered in a single-dish monitoring program. }
{We have used the Karl G. Jansky Very Large Array (JVLA) at the 1~cm band to map five highly-excited metastable inversion transitions  of NH$_3$, (J,K)=(6,6), (7,7), (9,9), (10,10), and (13,13), in W51~IRS2 with $\sim$0\pas2 angular resolution. }
{We present detections of both thermal (extended) ammonia emission in  the five inversion lines, with rotational states ranging in energy from about 400 to 1700~K, 
 and point-like ammonia maser emission in the (6,6), (7,7), and (9,9) lines.  
For the point-like emission, we estimate lower limits to the peak brightness temperatures of $1.7 \times 10^5$ K, $6 \times 10^3$ K, and $1 \times 10^4$ K 
for the  (6,6), (7,7), and (9,9) transitions, respectively, confirming their maser nature. 
The thermal ammonia emits around a Local Standard of  rest velocity of V$_{\rm LSR}$ = 60~\kms, near the cloud's systemic velocity,  
appears elongated in the East-West direction across 4\arcsec\,and is confined by the HII regions W51d (to the North), W51d1 (to the East), and W51d2 to the West.  
The \amm\,masers are observed in the eastern tip of the dense clump traced by thermal \amm,
 offset by 0."65 to the East  from its emission peak, and have a peak velocity at $\sim$47.5~\kms. 
No maser components are detected near the   systemic velocity.
The \amm\,masers arise close to but separated  (0\pas65 or 3500~AU) from the (rare)  vibrationally-excited SiO masers, 
which are excited in a powerful bipolar outflow driven by the deeply-embedded high-mass young stellar object (YSO) W51-North. 
This excludes that the two maser species are excited by the same object. 
Interestingly, the \amm\,masers originate at the same sky position as a peak in a submm line of SO$_2$ imaged with the SMA, 
tracing a face-on circumstellar disk/ring around W51-North.  
 In addition, the thermal emission from the most highly excited \amm\,lines, (10,10) and (13,13), 
 shows two main condensations, the dominant one towards W51-North  with the SiO/H$_2$O masers, 
and a  weaker peak at the \amm\,maser position. 
}
{  We propose a scenario where the ring seen in SO$_2$ emission is a circumbinary disk surrounding (at least) two high-mass YSOs, 
 W51-North (exciting the SiO masers) and a nearby companion (exciting the \amm\,masers), separated by 3500 AU. 
 This finding indicates a physical connection (in a binary) between the two rare SiO and \amm\,maser species.
 }

               
 \maketitle
           

\section{Introduction}

Intense NH$_3$ masers were first detected in the 1980s in non-metastable transitions from single-dish monitoring with the 100-m telescope at Effelsberg 
\citep{Wilson82, Madden86,Mauersberger87,WilsonHenkel88}. 
Recently, \citet{Henkel13} reported a total of 19 NH$_3$ maser lines towards the prominent massive star forming complex W51~IRS2,  
related to highly excited inversion doublets from both metastable and non-metastable levels. 
In addition to the  19 masers reported by \citet{Henkel13},  two additional  maser lines were previously reported  in W51-IRS2, raising the total number of \amm\,maser lines to 21. 
Most of these lines have no known counterpart in any other source 
(see  \citealt{Henkel13}, their Section 3.2, for detailed statistics).  
Therefore, W51~IRS2 is quite remarkable, providing the cosmic source with by far the largest number of detected  masing ammonia lines.

W51~IRS2 is part of W51 (at a distance of 5.4 kpc; \citealt{Xu09,Sato10}), one of the most luminous massive star forming  regions in the Galaxy (with $\sim10^6$ \ls), containing a cluster of O-type forming stars associated with three main regions: W51 IRS1, W51 Main, and W51 IRS2 itself.  
While  W51 IRS1 is associated with an evolved HII region (size$\sim$1 pc) and is deprived of molecular gas or dust, 
W51 Main  and W51 IRS2 (separated by approximately one arcminute) are sites of active massive star formation \citep[e.g.,][]{ZhangHo97}. 
W51 Main 
powers strong OH \citep{Etoka12} and \wat\, \citep{Imai02} masers near several ultracompact (UC) HII regions,  
which are associated with widespread \amm~thermal emission \citep{ZhangHo97}.  
 Interestingly, towards W51~IRS2, very strong maser emission is known from various species: 
 e.g., hydroxyl (OH), water (H$_2$O), and even silicon monoxide (SiO)  
 \citep[e.g.,][]{GaumeMutel87,Imai02,Eisner02}. 
In fact, W51~IRS2 is one of only three high-mass star forming regions known to exhibit SiO maser emission \citep{Hasegawa1986,Zapata09b},  
which originates from vibrationally excited levels  higher than 1000 K above the ground state and thus requires highly excited gas, 
possibly heated either by shocks or  radiation from a nearby protostar \citep[e.g.][]{Goddi09,Matthews10}.  
Therefore, W51~IRS2 is special because it powers the rare \amm\,masers as well as the rare SiO masers.

\begin{figure}
\includegraphics[angle=-90,width=0.5\textwidth]{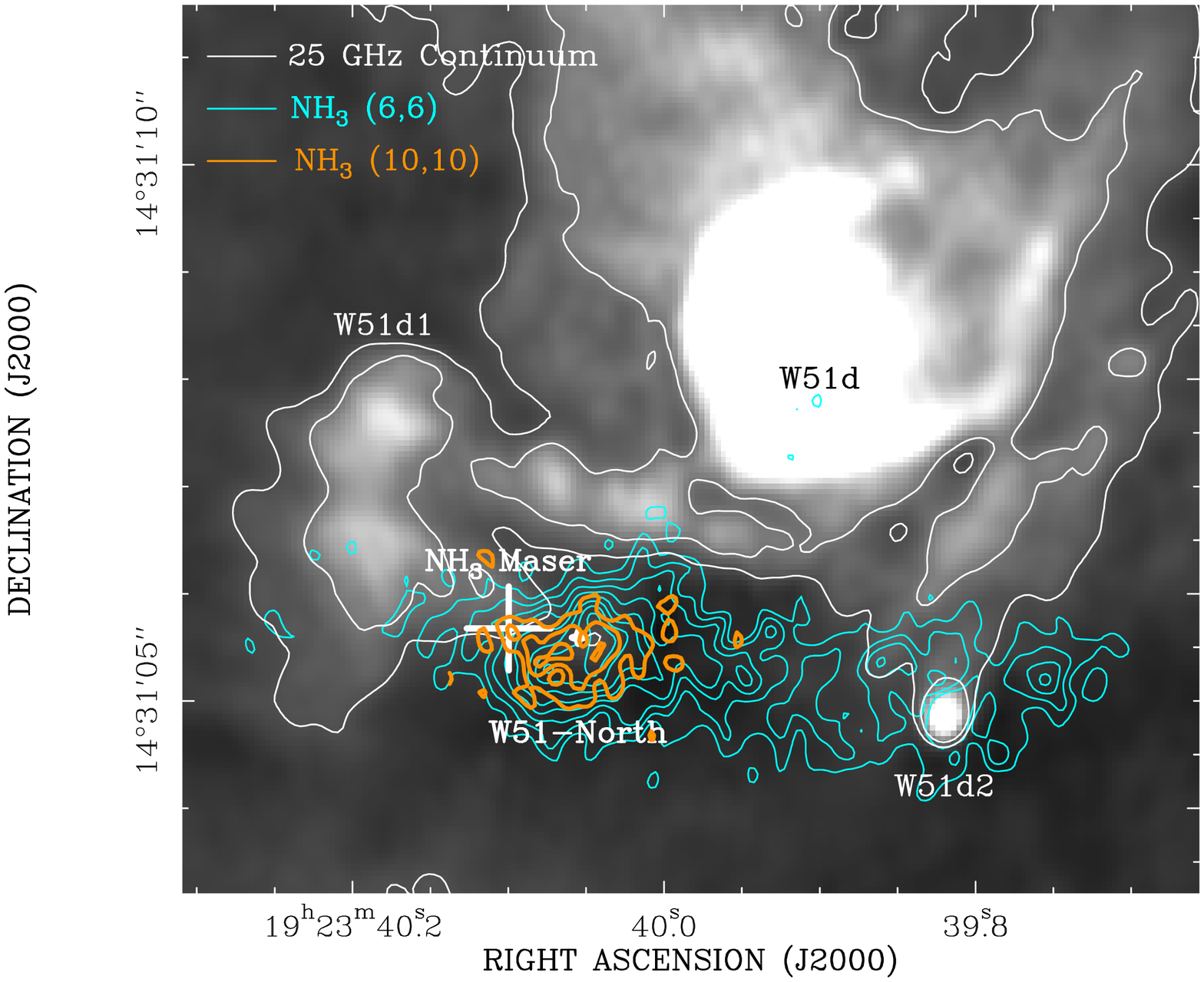}
\includegraphics[angle=-90,width=0.5\textwidth]{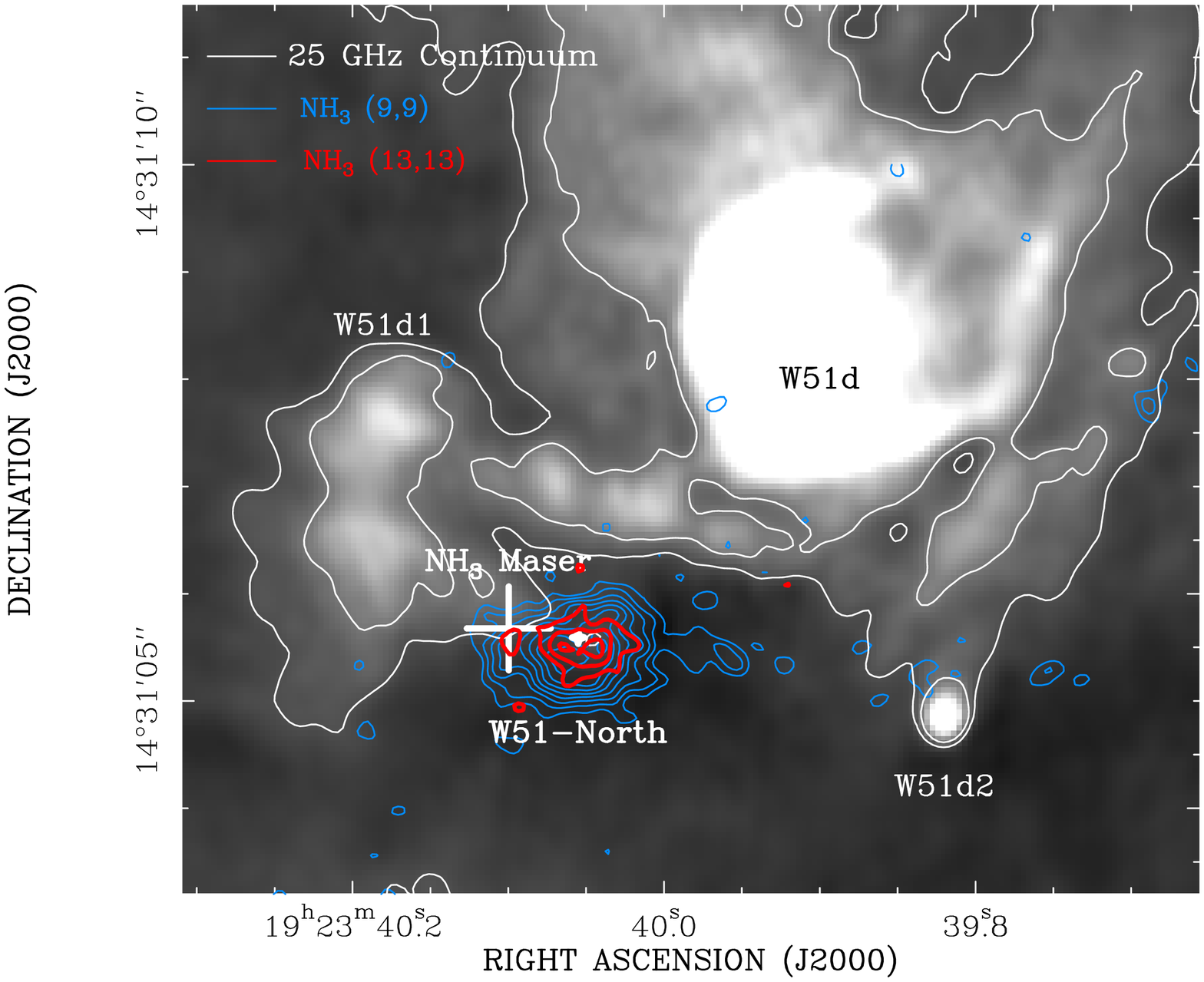}
\caption{
Overlay of the 25 GHz continuum emission (grey scale and white contours) 
 and the 0$^{th}$ moment images of the \amm\,(6,6), (9,9), (10,10), and (13,13) thermal emission lines 
 (cyan, blue, orange, and red contours, respectively) imaged with the JVLA in  W51~IRS2. 
 The 25 GHz continuum contours indicate 1 and 3 mJy flux levels per beam, respectively  
 (the peak is $\sim$160 mJy beam$^{-1}$ and the RMS is 0.4  mJy beam$^{-1}$). 
The \amm\,contours are 20\% to 100\% with steps of 10\% of the line peak for the (6,6) line (119 mJy beam$^{-1}$ km s$^{-1}$)  
and the (9,9) line (63 mJy beam$^{-1}$ km s$^{-1}$), 
and  30\%, 60\%, 90\% of the line peak for the (10,10)  (66 mJy beam$^{-1}$ km s$^{-1}$)  
and the (13,13) lines (65 mJy beam$^{-1}$ km s$^{-1}$). 
The intensity was integrated in the velocity range from 50.3 to 70.7 km s$^{-1}$ for the (6,6) line, 
 +50.4 to +66.0 km s$^{-1}$ for the (9,9) line, 
 +50.2 to +66.2 km s$^{-1}$ for the (10,10) line, 
and +50.0 to +66.8 km s$^{-1}$ for the (13,13) line, respectively. 
 The big white plus marks the position of the (6,6), (7,7), and (9,9) \amm\,masers.  
The small white plus marks the position of the W51-North protostar as inferred from SiO masers \citep{Eisner02}.  
Note that the most highly excited lines, the (10,10) and (13,13) transitions, show a peak
at the \amm\,maser position. The images displayed are not corrected for primary beam effects. 
}
\label{nh3+cont}
\end{figure}

While the Effelsberg monitoring spanning two decades enabled detection and study of the spectral profiles of these ammonia lines, 
their physical/spatial origin is still unknown. 
This is particularly relevant in the case of W51~IRS 2, since this star forming complex includes several centers of activity, within a few arcseconds (see Figure~\ref{nh3+cont}): 
a prominent hot core in a pre-HII stage  exhibiting  very rich chemistry and driving a powerful outflow (W51-North), 
a  UC-HII region exhibiting strong emission from many complex molecules (W51d2),
a compact HII region (W51d1) and a cometary HII region  (W51d) with little molecular gas associated \citep[e.g.,][]{Gaume93,Zhang98,Zapata10}.  
Interferometric mapping is therefore necessary to establish the origin of the NH$_3$ masers within W51~IRS 2.

\citet{Henkel13} identified three families of \amm\,maser lines, with emission near 57~\kms, 54.5~\kms, and 45~\kms,  respectively. 
The first two families of maser lines arise from non-metastable levels and are known for thirty years.  
 The third family was  not detected in the observations of   \citet{Mauersberger86,Mauersberger87},
but now shows maser emission from both metastable and non-metastable lines, well separated from the quasi-thermal emission near 60~\kms. 
\citet{Henkel13} suggested a possible correlation between this third family of  highly excited  \amm\,masers 
and vibrationally-excited SiO masers,  
 since these transitions require particularly high excitation conditions, 
  and both maser species show a velocity component around 45~\kms\,with a secular velocity drift of  $\sim$0.2~\kmsy.  
Interferometric observations are clearly required to confirm (or refute) a physical relation between these two rare maser species. 


Here, we present interferometric observations of five metastable inversion lines from \amm: (J,K) = (6,6), (7,7), (9,9), (10,10), and (13,13). 
Recent interferometric work has focused on non-metastable lines (e.g., \citealt{Walsh07} and \citealt{Hoffman14}, and references therein). 
Interferometric observations of metastable ammonia masers above the (3,3) level are instead rare 
( one case is the (6,6) maser in NGC6334 by \citealt{Beuther07}). 
Metastable \amm\,masers are particularly interesting because,  in contrast to non-metastable masers (or even to other ubiquitous maser species, like \wat), 
thermal emission from the same transition is also found in the environment of the maser activity. 
Therefore, these provide the unique chance to investigate maser excitation and its relation with star formation activity. 

 This paper is the second in a series of \amm\,multilevel imaging studies in high-mass star forming regions 
(for the first, see Goddi et al., 2014, hereafter Paper I).  
The current paper is structured as follows.
The  observational setup and data calibration procedures are described in \S 2.
Spectral profiles and  images of different maser transitions are presented in \S 3. 
In \S 4, we discuss the relation of the \amm\,masers to the star formation activity in W51~IRS2  (\S 4.1)  
and potential pumping mechanisms of these \amm\,masers (\S 4.2). 
 Finally, 
 we present a Summary in \S 5. 
 
\begin{table*}
\caption{Parameters of JVLA observations toward W51.}             
\label{obs}      
\centering                        
\begin{tabular}{ccccccc} 
\hline\hline                 
\noalign{\smallskip}
\multicolumn{1}{c}{Transition$^{a}$} & \multicolumn{1}{c}{$\nu_{\rm rest}$} & \multicolumn{1}{c}{$E_u/k^{b}$} & \multicolumn{1}{c}{Date}  & \multicolumn{1}{c}{JVLA} & \multicolumn{1}{c}{Beamwidth$^{c}$}  & \multicolumn{1}{c}{RMS$^{d}$} \\ 
\multicolumn{1}{c}{(J,K)} & \multicolumn{1}{c}{(MHz)} & \multicolumn{1}{c}{(K)} & \multicolumn{1}{c}{(yyyy/mm/dd)} & \multicolumn{1}{c}{Receiver}  & \multicolumn{1}{c}{$\theta_M('') \times \theta_m(''); \ P.A.(^{\circ})$} &  \multicolumn{1}{c}{(mJy/beam)}   \\
\noalign{\smallskip}
\hline
\noalign{\bigskip}
(6,6)   & 25055.96    &  408  & 2012/05/31 & K  & $0.29 \times 0.23;\ -21$  & 1.3 \\
(7,7)   & 25715.14    &  539  & 2012/05/31 & K  & $0.28 \times 0.23;\ -2$  & 1.4 \\
(9,9)   & 27477.94    &  853  & 2012/06/21 & Ka & $0.24 \times 0.22;\ +53$  & 1.3 \\
(10,10) & 28604.75    & 1035  & 2012/08/07 & Ka & $0.23 \times 0.21;\ +61$  & 2.9 \\
(13,13) & 33156.84    & 1691  & 2012/06/21 & Ka & $0.20 \times 0.18;\ +52$  & 1.9 \\
\noalign{\smallskip}
\hline   
\end{tabular}
\tablefoot{\\
(a) Transitions include ortho-\amm~($K=3n$) and para-\amm~($K\neq3n$).  \\
(b) Energy above the ground  reported from the JPL database.  \\ 
(c) Synthesized beams  in images made with the CASA task CLEAN with a robust parameter set to 0.5. \\
(d) RMS noise in a $\sim$0.4~\kms~channel without primary beam correction. 
After primary beam correction,  the noise level increases by up to 25\%.  
}
\end{table*}


\section{Observations}
\label{observations}
Observations of NH$_3$ towards the W51 complex were conducted using the Karl G. Jansky Very Large Array (JVLA) of the National
Radio Astronomy
Observatory (NRAO)\footnote{NRAO is a facility of the National Science Foundation operated under cooperative agreement by Associated Universities, Inc.} in the B configuration.
By using the broadband JVLA K- and Ka-band receivers, we
observed  a total of five metastable inversion transitions of NH$_3$: 
($J,K$)=(6,6), (7,7), (9,9), (10,10), and (13,13)  at  the 1~cm band, with frequencies going from \ $\approx$25.1~GHz for the (6,6) line to $\approx$33.2~GHz for the (13,13) line.    
Transitions were observed in pairs of (independently tunable) basebands 
during 6h tracks (two targets per track:  W51 -- this paper; NGC7538~IRS1 -- Paper I) on three different dates in 2012: 
the (6,6) and (7,7) lines on May 31 at K-band, the (9,9) and (13,13) lines on June 21, 
and the (10,10) transition on August 7, both  at Ka-band.  
Each baseband had eight sub-bands with a 4~MHz bandwidth  each ($\approx$40~\kms\ at 30~GHz), providing a total coverage of 32~MHz ($\approx$320~\kms\ at 30~GHz). 
Each sub-band consisted of 128 channels with a separation of 31.25~kHz ($\approx$0.3~\kms\ at 30~GHz). 
The typical on-source integration time was about 80 min. Each transition was observed with 
``fast switching'', where 80s scans on-target were alternated with 40s
scans  on the nearby (1.2$^{\circ}$ on the sky) QSO J1924+1540  (measured flux density 0.6--0.7~Jy, depending on frequency). 
We derived  absolute flux calibration from observations of 3C~48 ($S_{\nu}$ = 0.5--0.7~Jy, depending on frequency), 
and bandpass calibration from observations of 3C~84 ($S_{\nu}$ = 27--29~Jy, depending on frequency).

The data were edited, calibrated, and imaged in a standard fashion using the Common Astronomy Software Applications (CASA) package. 
We fitted and subtracted continuum emission  from the spectral line data in the uv plane  
using CASA task UVCONTSUB, combining the continuum (line-free) signal from all eight sub-bands around the \amm\ line. 
Before imaging the \amm\ lines, we performed self-calibration on the strong  (6,6) \amm\,maser at 47.6 \kms\,(peak flux density $\sim$5 Jy). 
We then applied the self-calibration solutions from the reference channel to the dataset containing the lines (6,6) and (7,7). 
Since the (9,9) maser line  was much weaker than the (6.6) line (peak flux density $\sim$0.4 Jy), 
we did not perform self-calibration on the dataset containing the (9,9) and (13,13) lines (likewise for the 10,10 transition). 
 We discuss the accuracy in both absolute and relative astrometry among different transitions in \S~\ref{spots_maps}.  
Using the CASA task CLEAN, we imaged the W51~IRS2 region with a cell size of 0\farcs04, covering a 
20\arcsec~field around the position:  $\alpha(J2000) = 19^h 23^m 40^s.00$,   $\delta(J2000) = +14^{\circ} 31' 06$\pas0. 
We adopted Briggs weighting with a ROBUST parameter set to 0.5 and smoothed the velocity resolution to 0.4~\kms, for all transitions. 
The resulting synthesized clean beam FWHM was \ 0\farcs19--0\farcs26 \ 
and the typical RMS noise level per  channel was \ $\approx$1.5~mJy~beam$^{-1}$. 
The observations were conducted pointing the telescopes at W5-IRS1, with a sky position of 
 $\alpha(J2000) = 19^h 23^m 42^s.00$,   $\delta(J2000) = +14^{\circ} 30' 50$\pas0   
(that way both W51-North and W51-South were included in the JVLA antennas' primary beam). 
Therefore, we applied primary beam corrections during cleaning (of the order of 15-25\%, depending on transition). 
Table~\ref{obs} summarizes the observations.

\begin{table}
\caption{Parameters of the \amm\, maser lines observed toward W51-North.}             
\label{lines}      
\centering                        
\begin{tabular}{cccccccccccccl} 
\hline\hline                 
\noalign{\smallskip}
\multicolumn{1}{c}{Line} &  & \multicolumn{1}{c}{F$_{\rm peak}$} &  \multicolumn{1}{c}{V$_{c}$} & \multicolumn{1}{c}{$\Delta V_{1/2}$}&\multicolumn{1}{c}{F$_{\rm int}$} &\multicolumn{1}{c}{T$_b$} \\ 
\multicolumn{1}{c}{(J,K)} &  &  \multicolumn{1}{c}{(Jy)}  & \multicolumn{1}{c}{(km/s)}     & \multicolumn{1}{c}{(km/s)}&  \multicolumn{1}{c}{(Jy~km/s)} &  \multicolumn{1}{c}{($10^4$~K)}\\
\noalign{\smallskip}
\hline
\noalign{\smallskip}
(6,6)   &     &  4.88 & 47.6 & 1.4 & 7.08 & 17.2\\
(7,7)   &     &  0.20 & 47.1 & 1.2 & 0.26 & 0.7 \\
        &     &  0.17 & 49.7 & 1.0 & 0.19 & 0.6 \\
(9,9)   &     &  0.35 & 47.4 & 1.5 & 0.52 & 1.0 \\
\noalign{\smallskip}
\hline   
\end{tabular}
\tablefoot{
The peak fluxes (F$_{\rm peak}$, col. 2), the central velocities (V$_{c}$; col. 3), the FWHM line-width ($\Delta V_{1/2}$; col. 4), and the velocity-integrated flux (F$_{\rm int}$; col. 5) are estimated from single-Gaussian fits to the spectral profiles in Figure~\ref{spec_nh3_mas}. 
 The brightness temperatures T$_b$ are estimated using a beamsize of 0\pas25 (upper limit to the unresolved maser size), 
 and should be regarded as lower limits. 
 The velocities reported  here (and throughout the paper) are Local Standard of Rest (LSR) velocities.
}
\label{nh3_lines}
\end{table}

\section{Results}
\label{res}
We have mapped five highly-excited metastable inversion transitions  of NH$_3$ (J,K)=(6,6), (7,7), (9,9), (10,10), and (13,13) in W51~IRS2, with $\sim$0\pas2 angular resolution 
We detected both thermal (extended) ammonia emission in five inversion lines, with rotational states ranging in energy from about 400 to 1700~K, 
 and point-like ammonia maser emission in the (6,6), (7,7), and (9,9) lines (Table~\ref{nh3_lines}). 
Using the line-free subbands, we also produced images of the radio continuum emission with an RMS noise level of 0.4~mJy~beam$^{-1}$, at the frequencies corresponding to the inversion lines (25-33.2 GHz).  

In the following we will discuss
the spatial and velocity distribution of the molecular gas  estimated from the thermal \amm\,emission
with respect to  the ionized gas traced by the continuum emission (Sect.~\ref{thermal}), 
spectral profiles of the maser and thermal ammonia lines (Sect.~\ref{spec}), 
and maps of ammonia maser spots (Sect.~\ref{spots_maps}). 

\subsection{Distribution of thermal ionized (continuum) and molecular (\amm) gas}
\label{thermal}
In Figure~\ref{nh3+cont}, we present the integrated thermal emission of the (6,6), (9,9), (10,10), and (13,13) lines overlaid on the 25 GHz continuum map\footnote{The distribution of the (7,7) line emission is similar to that of the (6,6) line, but the signal-to-noise ratios were lower, therefore we do not display it.};  
the position of the \amm\,masers is indicated by a white cross. 
At 25 GHz, we detected strong free-free continuum emission arising from the HII regions associated with several high-mass YSOs forming in the W51~IRS2 cluster. 
The most prominent ones are the extended cometary HII region W51d and the two compact HII regions W51d1 and W51d2 (see Table~\ref{ysos}). 
This emission has already been mapped at different (centimeter) wavelengths with the VLA at different resolutions \citep[e.g.,][]{Gaume93}. 
The dense molecular gas traced by the \amm\,thermal emission is observed at the southern edge of the W51d HII region,   
and is elongated along the East-West direction across 4\arcsec.  
Apparently, it is confined by the ionized gas: W51d1 to the East, W51d2 to the West, and W51d to the North. 
Among the transitions observed by us, the extended distribution of the thermal emission is 
 best traced by the lower excitation (6,6) line, 
  which, interestingly, exhibits a clumpy structure along a belt of emission from W51-North to W51-d2. 
Figure~\ref{nh3+cont} shows that the \amm\,masers are enclosed within the lowest contour of the
thermal \amm\,emission,  
and are observed in the eastern tip of the dense clump traced by thermal \amm.   
It is interesting to note that the most highly excited lines, (10,10) and (13,13), 
show the strongest emission around W51-North, but also weaker emission near the \amm\,maser position.

\begin{figure}
\includegraphics[width=0.5\textwidth]{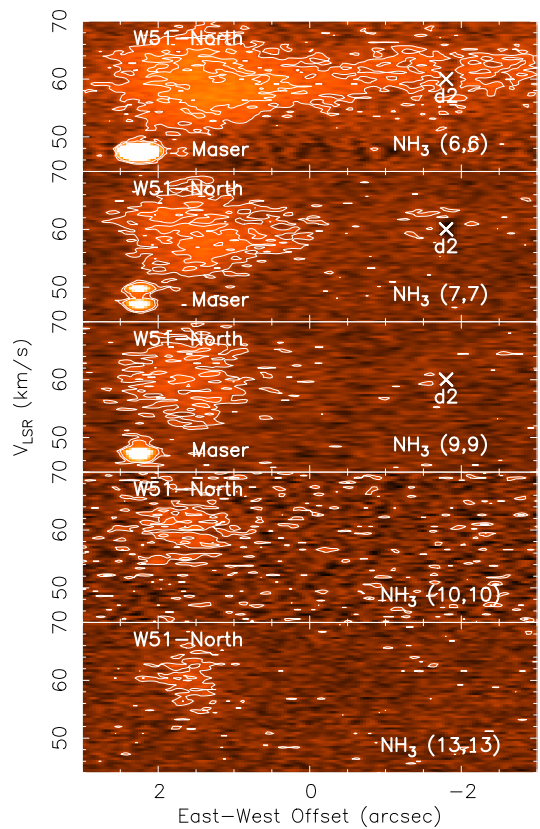}
\caption{
Position-velocity diagrams of the (J,K) $=$ (6,6), (7,7), (9,9), (10,10), and (13,13)  lines observed with the JVLA B-array towards W51~IRS2. 
The cut is taken at the peak of the \amm\,core and elongated East-West (i.e. P.A. $=$ 90\degree). 
 The zero offset in R.A.  corresponds to the phase-center position used in imaging:  $\alpha(J2000) = 19^h 23^m 40^s.00$,   $\delta(J2000) = +14^{\circ} 31' 06$\pas0.   
The contours are 1.5 and 3 mJy beam$^{-1}$ for the (6,6) line, and 1 and 2 mJy beam$^{-1}$ for the remaining lines.  
The colors are linear from 0 to 10  mJy beam$^{-1}$.  
The images were constructed with a 0\pas04 pixel for all transitions. 
The weak absorption in W51d2 is also indicated with a white cross. 
Note that the images displayed are not corrected for primary beam effects. 
}
\label{pv}
\end{figure}

In Figure~\ref{pv}, we also present the position-velocity (pv) diagrams of the (6,6), (7,7), (9,9), (10,10, and (13,13) lines. 
While the (6,6) line shows the most extended emission (from the maser to d2), the more highly excited lines show strong emission 
only in the hot-core excited by W51-North \citep[e.g.,][]{Zapata09}. 
From the pv-diagrams,  it becomes  evident that the \amm\,masers are separated in both velocity and position  
 from the peak of thermal \amm\,emission tracing the dense gas in W51-North. 
The thermal \amm\,emits in a velocity range from 50 to 71 km s$^{-1}$ and peaks at around 60~\kms\,(the cloud's systemic velocity), 
whereas the three masers peak at 47--50~\kms\,(see Table~\ref{nh3_lines}). 

\subsection{Spectral profiles of the \amm\,masers and thermal emission}
\label{spec}
\begin{figure}
\includegraphics[width=0.45\textwidth]{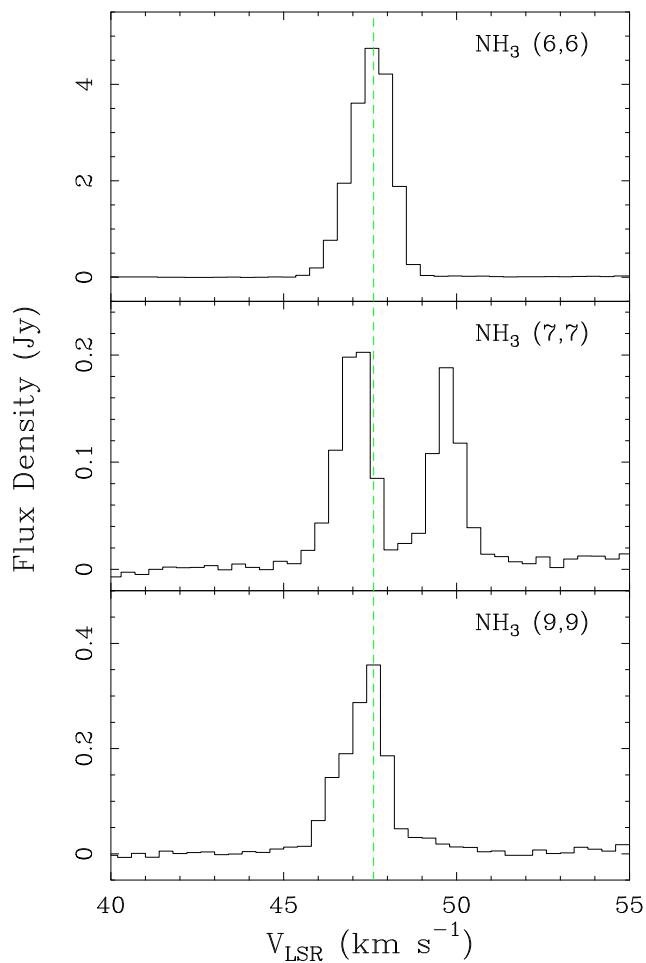}
\caption{Spectral profiles of the maser emission from the metastable NH$_3$ (6,6), (7,7), (9,9) inversion transition  lines observed toward W51-North with the JVLA B-array. 
The radial velocities are derived with respect to the LSR. 
The vertical dashed line indicates a velocity of 47.6~\kms. 
The velocity resolution is 0.4~\kms. 
Note that different flux density scales are adopted for different transitions.
Note also that the two ortho transitions, the (6,6) and (9,9) lines, show similar profiles, 
whereas the only para transition, the (7,7), shows a double-peaked profile  (see text). 
}
\label{spec_nh3_mas}
\end{figure}

For each transition, we produced spectra by mapping each spectral channel and summing the flux density in each channel map towards the compact maser emission (Figures~\ref{spec_nh3_mas}).  
The (6,6) and (9,9)  ortho-transitions show similar central velocities  (with respect to the local standard of rest or LSR), 
V$_c$=47.6, 47.4~\kms\,(with a scatter smaller than the velocity resolution of 0.4~\kms)
and velocity widths, $\Delta V_{1/2}\sim1.5$~\kms, as determined from single-Gaussian fits (Table~\ref{nh3_lines}). 
The observed line-widths are relatively large for maser emission ($>1$~\kms) and the line profiles appear rather asymmetric. 
Both elements may result from blending of different maser features. 
The presence of multiple components can be confirmed for the (7,7) transition, which shows a double-peaked profile, 
with two narrower components ($\Delta V_{1/2}\sim1$~\kms) at 47.1 and 49.7~\kms, respectively. 
 The (7,7) line was the only one out of 19 maser lines observed with Effelsberg to  show two velocity components below 50~\kms\,\citep{Henkel13}, indicating that this is a unique tracer.
Based on our JVLA imaging, we can now assess that the two components are separated not only in velocity but also spatially (see \S~\ref{spots_maps}). 

The inversion lines that we imaged with the JVLA are part of a monitoring study with the Effelsberg 100-m telescope \citep{Henkel13}. 
Therefore, we can compare our interferometric measurements with single-dish observations.  
The monitoring study revealed a velocity drift of 0.2~\kms~yr$^{-1}$ and a significant flux variation in different \amm\,maser lines. 
Therefore, we compare our results with the nearest single-dish observations in time.  
The (6,6) and (7,7) lines were observed with the 100-m telescope in April 2012, just two months before our JVLA observations. 
\citet{Henkel13} report:
V$_c$=47.2~\kms, $\Delta V_{1/2}\sim1.5$~\kms, and $F_{\rm peak}$=5.3 Jy for the (6,6) line;  
V$_c$=47.3 and 49.9~\kms, $\Delta V_{1/2}\sim1.4$ and 1.2~\kms, and $F_{\rm peak}$=0.2 and 0.18 Jy for the two velocity components of the (7,7) transitions, respectively. 
 This is in excellent agreement with our findings (see Table~\ref{nh3_lines}). 
The (9,9) line was observed with the 100-m telescope in August 2011, one year before our JVLA observations, 
and had V$_c$=46.8~\kms, $\Delta V_{1/2}\sim1.2$~\kms, and $F_{\rm peak}$=0.07 Jy. 
Taking into account the secular velocity drift and allowing for maser emission flux variation,
  our interferometric measurement of the (9,9) transition is also consistent with the single-dish observations. 

  It is worth noting that the excellent agreement between the integrated fluxes of the (6,6) and (7,7) lines measured with both the single-dish and the interferometer, 
   implies that there is no significant extended maser emission that might have been missed by the JVLA. 
  Therefore, we are confident that the maser emission comes from a region smaller than 0\pas2.  


The maser nature of the \amm\,emission from these highly excited lines has been inferred from the single-dish monitoring with Effelsberg, based on emission variability. 
Our interferometric measurements now allow us to calculate peak brightness temperatures, $T_b$, based on equation : 
$T_b= S (c/\nu)^2/(2k_Bd\Omega)$,
where $S$ is the peak flux density, $\nu$ the line frequency and $d\Omega$ the solid angle of the source.
The compact \amm\,emission is not resolved  in our JVLA images, therefore we can only give lower limits to the true brightness temperature. 
Taking a representative beamsize of 0\pas25 as an upper limit to the maser size, given the observed line intensities, 
we  calculated   lower limits to the peak brightness temperatures of  $1.7 \times 10^5$ K, $6 \times 10^3$ K, and $1 \times 10^4$ K,  
for the  (6,6), (7,7), and (9,9) transitions, respectively. 
If the emission were thermal, it would require a local exciting source with a temperature in excess of these values which we consider extremely unlikely. 
 Therefore, we conclude that our interferometric measurements confirm the maser nature for the emission coming from these highly-excited \amm\,metastable lines.


\begin{figure}
\includegraphics[width=0.45\textwidth]{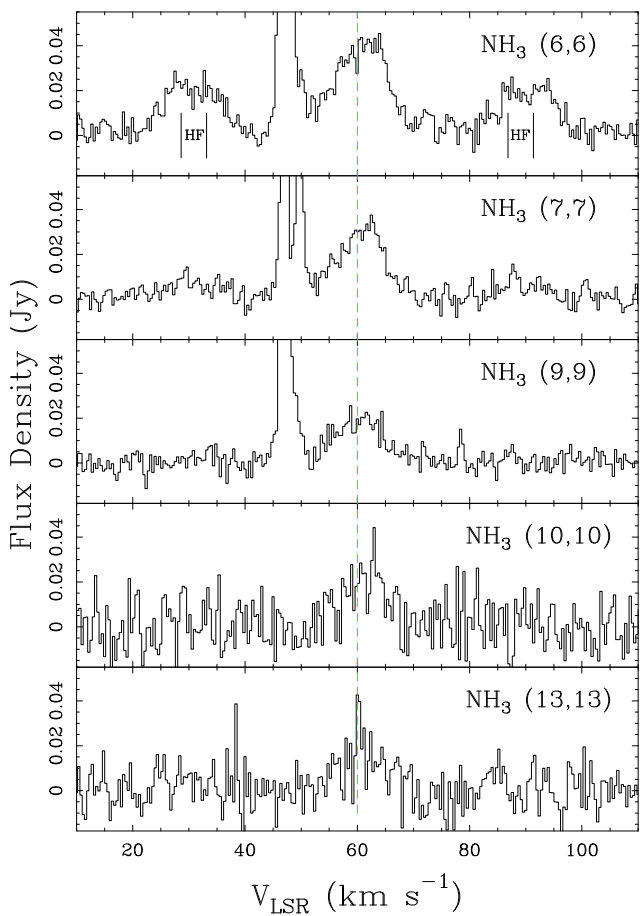}
\caption{
Spectral profiles of the thermal  (and maser) emission from the metastable NH$_3$ (6,6), (7,7), (9,9), (10,10), and (13,13) inversion transitions,    integrated in a region with size of 0\pas8 around the maser position. 
 The spectra shown here are the same as in Figure~\ref{spec_nh3_mas}, but with a narrower flux density and a wider velocity range to highlight the weaker thermal emission, 
The vertical dashed line indicates a velocity of 60~\kms. 
The velocity resolution is 0.4~\kms. 
Note that the hyperfine satellite lines (separated by $\sim \pm27-31$ \kms) are clearly detected for the (6,6) line 
 (indicated with black vertical lines),   
              but they are also still visible in the (7,7) spectrum.
             The upper state energy levels of transitions shown here are $\sim 408-1691$\,K (see Table~\ref{obs}).
}
\label{spec_nh3_therm}
\end{figure}

As discussed in \S~\ref{thermal}, besides maser emission, in W51~IRS2 we also observe extended thermal emission from \amm.  
In Figure~\ref{spec_nh3_therm} we show the spectral profiles of the \amm\,thermal emission at the maser position\footnote{The spectra displayed in Figure~\ref{spec_nh3_therm} 
 are integrated in a region with size of 0\pas8 around the maser position; therefore, they do not represent the total thermal emission, which extends across a few arcseconds.}.  
As already pointed out, the thermal emission towards the compact maser has a different velocity with respect to the maser, 
around 60~\kms, i.e. the cloud's systemic velocity. 
Remarkably, we find prominent hyperfine  satellite lines in the lowest-excitation (6,6) line, 
 spaced $\sim$27-31 \kms\, from the main line\footnote{Hyperfine satellite emission is also visible for the (7,7) doublet, but we refrain from estimating an optical depth, owing to low SNR.} (Figures~\ref{spec_nh3_therm}, top panel).
In fact, owing to interaction with the quadrupole moment of the nitrogen nucleus, 
each \amm\,inversion line is split into five components, 
a ``main component'' and four symmetrically spaced ``satellites'' with nearly equal intensities (at least in LTE),  
which makes up the quadrupole hyperfine structure (HFS). 
Since the relative line strengths of the satellite lines are a very small fraction of the main line intensity ($<$1\% for the transitions targeted here), 
the hyperfine satellites are  usually optically thin. 
The mere presence of measurable hyperfine satellites in the profile of  the (6,6) line however indicates very large optical depths 
also in these highly excited transitions.
We estimate an optical depth for the main hyperfine component of the (6,6) line of order of 100.  
\citet{Mauersberger86,Mauersberger87} present a single-dish multilevel study of ammonia from both metastable and non-metastable levels in W51~IRS2, and  find that most of the targeted transitions are very optically thick ($\tau>10$), with the (8,6) doublet showing the largest optical depth of 100, consistent with the value determined here for the (6,6) doublet. 

\subsection{Velocity-channel centroid maps of \amm\,maser emission}
\label{spots_maps}
\begin{figure*}
\centering
\includegraphics[angle= -90,width=\textwidth]{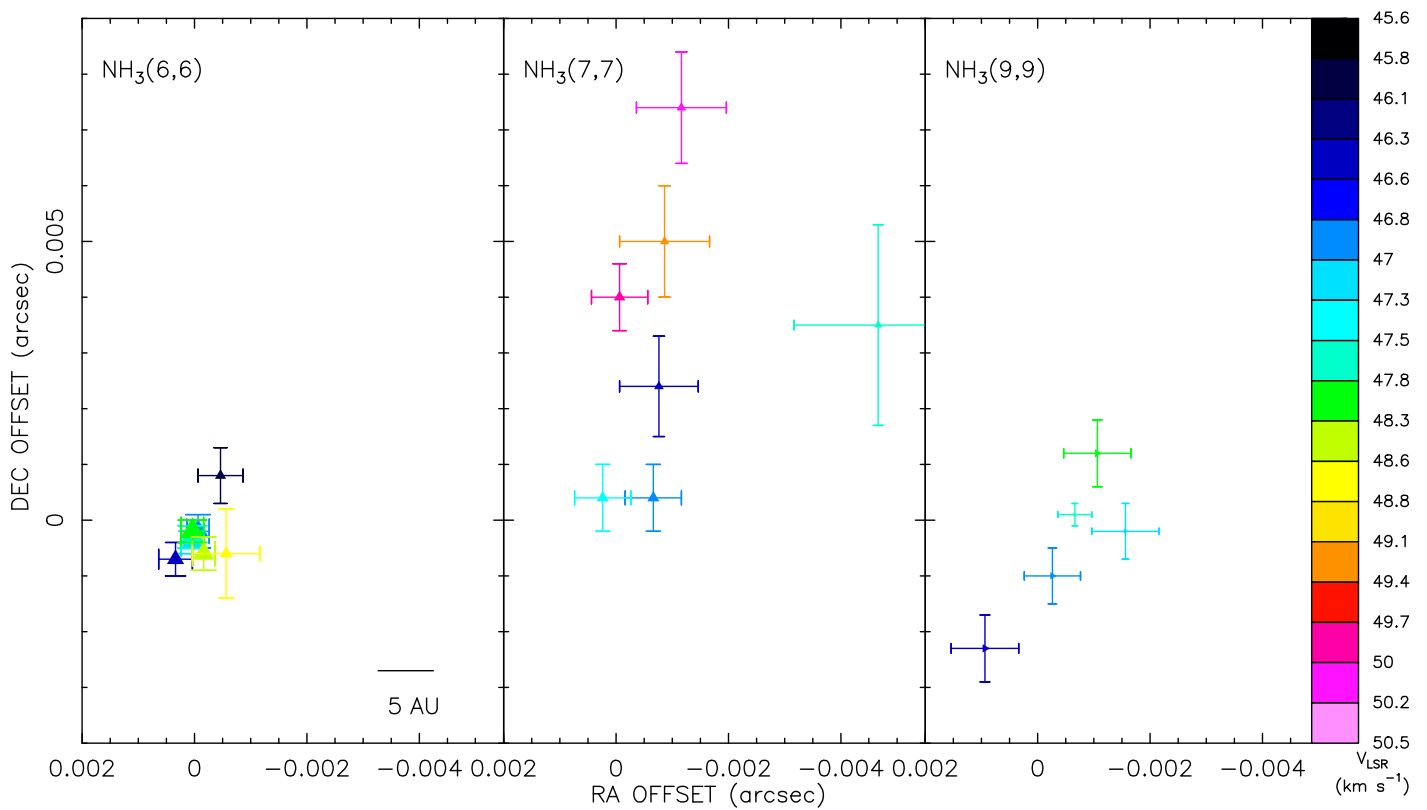}
\caption{Distribution of \amm\,maser emission in W51-North imaged with the JVLA.
Emission centroids of the (6,6), (7,7) and (9,9) lines  ({\it filled triangles}) 
are shown as a function of velocity in the left, middle, and right panel, respectively. 
 {\em Color} denotes  L.O.S.  velocity (color scale on the right-hand side). 
The size of the symbols  scales logarithmically with the flux density.  
The size of the crosses represents the $1\sigma$ formal fitting error to the position of the emission in individual velocity channels. 
 The positional offsets are from  
$\alpha(J2000) = 19^h 23^m 40\rlap{.}^s0997$,   $\delta(J2000) = 14^{\circ} 31' 05$\pas677. 
The linear spatial scale is given in the bottom right corner of the left panel. 
}
\label{spots}
\end{figure*}

 In each velocity channel, the maser emission is unresolved in all transitions. 
We fitted a two-dimensional ellipsoidal Gaussian model to each unresolved emission component in a $50 \times 50$ pixel   area of each channel map. 
Figure~\ref{spots} shows the positions (and associated errors), radial velocities, 
and intensities of the peak emission in different velocity channels for each inversion line.
 One striking feature shown in these maps is that the centroids of the maser spot distributions are coincident 
(within a few mas) for the three \amm\,transitions (see Table~\ref{mas_abs_pos} for their absolute positions).  
Another striking feature is that, while the distribution of maser spots in the (6,6) and (9,9)  ortho-transitions is fairly compact 
(within 1-2 mas),  
the emission from the para (7,7) line shows a distribution elongated North-South (across 8 mas), with also a separation in LSR velocities.  
 The blue-shifted velocities originate from the South while the red-shifted velocities are found in the North. 
This spatial and velocity distribution may explain the double-peaked spectral profile shown in Figure~\ref{spec_nh3_mas} (middle panel). 

In order to establish the accuracy of the superposition of different maser features from different maser lines shown in Figure~\ref{spots},
in the following we analyze different contributions to their positional errors.  

 The {\it relative} positional errors of maser features  from one specific maser transition are noise limited. 
 In particular, the positional uncertainty is proportional to the synthesized beamsize and is inversely proportional to the signal-to-noise ratio of the channel maps: 
 $\delta\theta = 0.5 \theta_B/{\rm SNR}$, 
 where $\theta_B$ is the FWHM of the synthesized beam, and 
 SNR is the peak intensity divided by the RMS noise in a particular velocity channel.
With typical SNR$>$100 and $\theta_B=$0\pas25, the positional accuracy is estimated to be of order of a fraction of mas 
(formal errors in the Gaussian fitting, $1\sigma=$0\pas0002-0\pas001, are consistent with this expectation). 
We thus conclude that the distribution of the emission in the three inversion lines, the (6,6) transition,
which is compact within 2 mas, the (7,7) transition, which extends across 8 mas, and 
the (9,9) transition, extending across 4 mas, 
 indicates different spatial structures and is not a consequence of relative positional errors.

When comparing the {\it absolute} positions of individual maser lines, two extra sources of uncertainty need to be taken into account. 
The (6,6) and (7,7) lines were observed simultaneously adopting two different basebands, having a $\sim$0.7~GHz separation. 
The accuracy in their relative separation is limited not only by SNR (and bandpass calibration), 
but potentially contains also a systematic position shift,  
which is proportional to the source offset with respect to the phase-tracking center used in correlation, 
multiplied by the difference in the rest frequencies between the two transitions 
(for a more detailed discussion see e.g. \citealt{Imai10}, \citealt{Goddi09}, and Appendix A in \citealt{Goddi11}). 
Considering that the maser source is located 30\arcsec\,north-west from the phase-tracking center (see \S~\ref{observations}), 
and the separation between the (6,6) and (7,7) maser lines is $\sim$0.7~GHz, 
this contribution can be potentially a significant fraction of an arcsecond. 
 Methods to remove  this potential positional offset  are described in \citet[][see their \S~2]{Goddi09},  in the case of SiO maser lines observed with the VLA. 
In order to test our relative astrometry, we also made images of the (7,7) masers without applying the self-calibration solutions from the (6,6) line. While the signal-to-noise ratio of the resulting  (7,7)  channel maps  was lower than the self-calibrated ones,  the relative position between the (6,6) and (7,7) maser emission was comparable in the two cases (with and without self-calibration applied), indicating that their  relative astrometry is correct within a few mas. 

Finally, the precision of the absolute position of interferometric images is limited by phase solutions. 
The (9,9) line was observed on a different date and a comparison with the other two lines requires absolute sky coordinates. 
Our astrometric measurements employ the fast switching phase-referencing technique with a nearby strong calibrator 
(see \S~\ref{observations}). 
The theoretical error in computing the absolute astrometry is proportional to the separation of the calibrator from the target 
and the positional accuracy of individual antennas, 
and inversely proportional to the baseline length. 
 Under good conditions of phase stability, accurate antenna positions ($\sim$1~cm), 
 a close calibrator ($\sim1^{\circ}$) with accurately known position ($\sim$1~mas), 
 and rapid switching ($\sim$1~min), 
 the accuracy is of order of 1-2\% of the synthesized beam. 
 We estimate therefore an accuracy of 2-4~mas for the absolute astrometry.  
 Consistently, the centroid emissions of the (6,6) and (9,9) lines coincide within about 1--2 mas.   

 Since both the (6,6) and (9,9) lines show, besides similar positions, also the same velocity profile and linewidth, 
we  suggest that the two ortho transitions are coincident in position and velocity. 
On the other hand, the different spatial (and velocity) distribution of the (7,7) line cannot be a consequence of phase-errors, 
as argued above, therefore we conclude that the ortho- and para-\amm\,masers show a different structure and velocity. 

\begin{table}
\caption{Absolute positions of the maser emission centroids  for the three \amm\,transitions.}             
\label{mas_abs_pos}      
\centering                        
\begin{tabular}{lccc} 
\hline\hline                 
\noalign{\smallskip}
\multicolumn{1}{c}{\amm\,Line} & &RA (J2000) & DEC (J2000) \\
 &  & \multicolumn{1}{c}{(h:m:s)} &  \multicolumn{1}{c}{($^\mathrm{o}$:':")}  \\
\noalign{\smallskip}
\hline
\noalign{\smallskip}  
(6,6)          & &19:23:40.09970 & 14.31.05.6767  \\
(7,7) Blue & & 19:23:40.09968 & 14.31.05.6781 \\
(7,7) Red & &19:23:40.09966 & 14.31.05.6825  \\ 
(9,9)         & &19:23:40.09967 & 14.31.05.6766  \\ 
\noalign{\smallskip}
\hline   
\end{tabular}
\end{table}

\begin{table*}
\caption{High-mass YSOs in W51~IRS2.}             
\label{ysos}      
\centering                        
\begin{tabular}{lcllccc} 
\hline\hline                 
\noalign{\smallskip}
\multicolumn{1}{c}{Source} & &RA (J2000) & DEC (J2000) & & \multicolumn{1}{c}{Tracer} \\
\multicolumn{1}{c}{Name} &  & \multicolumn{1}{c}{(h:m:s)} &  \multicolumn{1}{c}{($^\mathrm{o}$:':")} & & \\
\noalign{\smallskip}
\hline
\noalign{\smallskip}  
W51d  & & extended & extended & & Cometary HII\\
W51d1  & & 19:23:40.175 & 14:31:07.65 & & UC-HII\\
W51d2  & & 19:23:39.820 & 14:31:04.90 & & UC-HII\\
W51-North & & 19:23:40.055 & 14:31:05.59 & & SiO/H$_2$O masers \\ 
W51-North "Companion" & & 19:23:40.0997 & 14:31:05.677 & & NH$_3$ thermal/masers / SO$_2$  \\ 
\noalign{\smallskip}
\hline   
\end{tabular}
\end{table*}

\section{Discussion}

For the first time, we imaged high-excitation maser lines from metastable levels of ammonia up to 850 K above the ground in a high-mass star forming region. 
Being an entirely new tracer, there are still many open questions on the nature of these maser lines,  but the latter can potentially open up new frontiers in high-mass star formation studies.  
In particular, key questions to address are: 
Are the \amm\,masers excited in a particular stage of high-mass star formation? 
Are they associated with outflows, circumstellar accreting gas, or quiescent cores? 
Why does the bulk of \amm\, gas radiate thermal emission and maser emission only at a specific location? 
Why do thermal and maser emissions peak at different velocities?  
Is there a connection between rare \amm\,and SiO masers?
What is the pumping mechanism of these high-excitation maser lines from \amm? 
In the following, we will try to address some of these key questions. 

\subsection{Connections of the \amm\,masers with star formation activity in W51~IRS2}
\label{sf}
Our interferometric measurements have enabled us to locate precisely the highly-excited \amm\,emission (both maser and thermal) in the surroundings of W51-North. 
The latter is a highly obscured YSO, with no mid-IR counterpart \citep[e.g.,][]{Okamoto01} 
or free-free emission from a UC-HII region \citep[e.g.,][]{Gaume93,Eisner02}, 
but it exhibits strong thermal dust emission at (sub)millimeter wavelengths \citep[e.g.,][]{Zapata08,Zapata09} 
as well as molecular emission typical of a hot core at an excitation temperature of 100-300 K \citep[e.g.,][]{Mauersberger87,Zhang98,Zapata10}. 
In addition, this YSO powers strong H$_2$O, OH, and (exceptionally) SiO masers, 
all located in the center of this molecular and dusty structure \citep[e.g.,][]{Imai02,Eisner02}.

In a series of papers, Zapata and collaborators revealed that the obscured high-mass protostar located in W51-North
is associated with a dusty circumstellar disk, a hot gas molecular ring, and a collimated outflow. 
The simultaneous presence of these elements makes W51-North a compelling target for high-mass star formation studies. 

\begin{figure}
\includegraphics[angle=-90,width=0.5\textwidth]{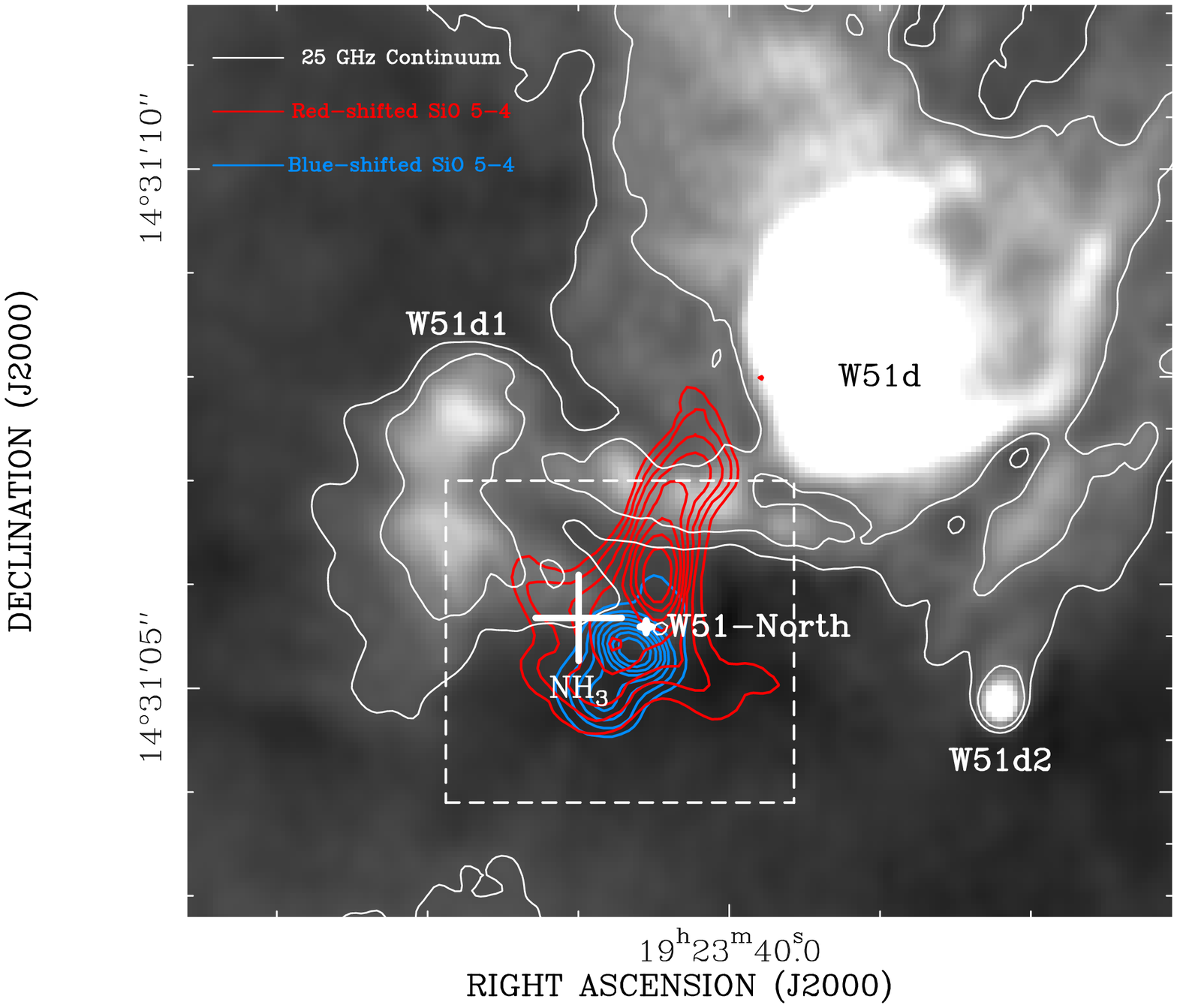}
\includegraphics[angle=-90,width=0.5\textwidth]{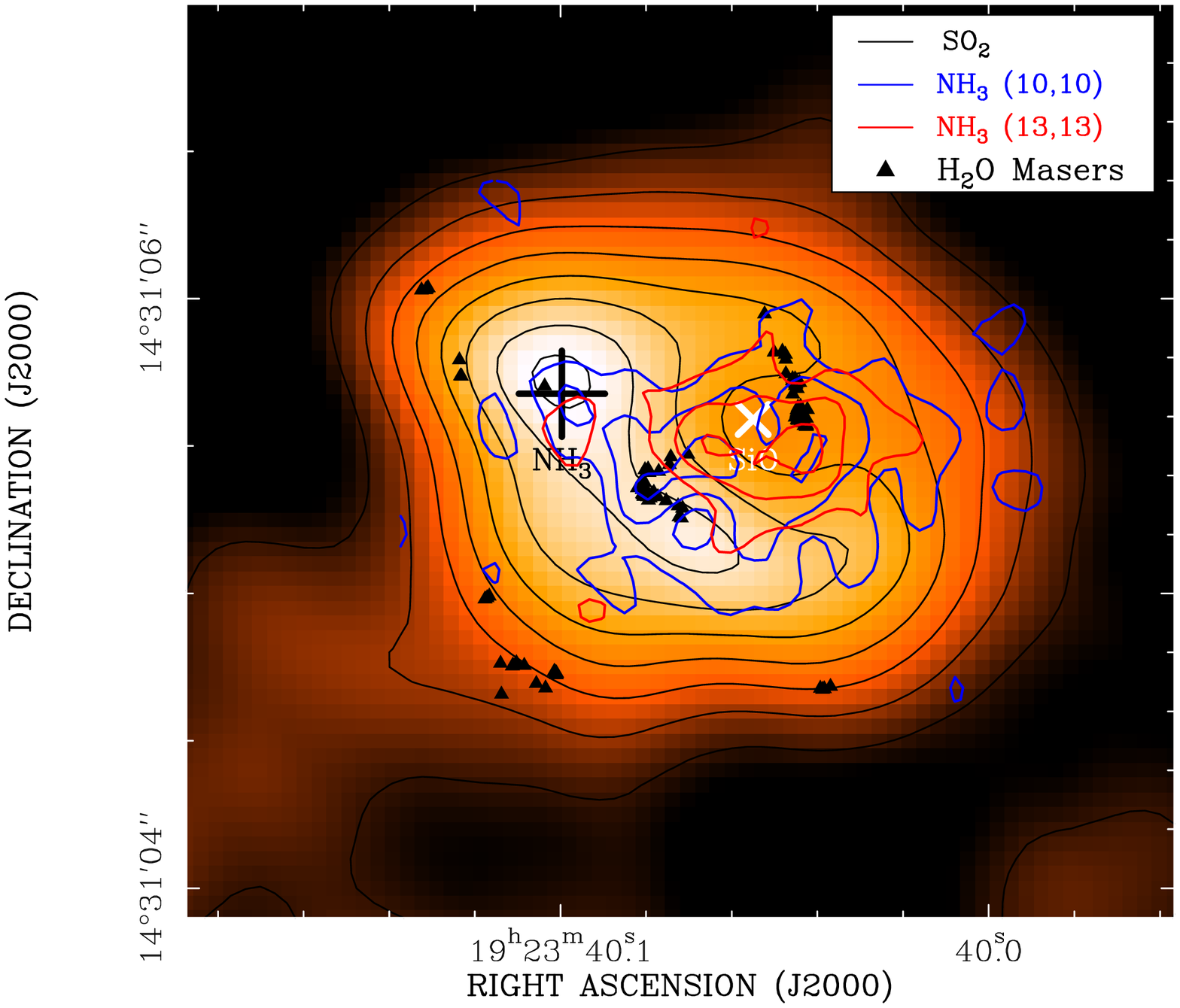}
\caption{ W51 IRS2. 
 {\it Upper Panel:} Overlay of the 25 GHz continuum emission imaged with the JVLA (grey scale and white contours) 
 and the SiO v = 0 J=5-4 $0^{th}$ moment image from the SMA (blue and red contours; \citealt{Zapata09}). 
 The 25 GHz continuum contours indicate 1 and 3 mJy flux levels per beam, respectively 
 (the peak is $\sim$160 mJy beam$^{-1}$ and the RMS is 0.4  mJy beam$^{-1}$). 
The SiO 5-4 line contours are from 20\% to 90\% with steps of 10\% of the line peak  
(5230 Jy beam$^{-1}$ km s$^{-1}$ for the red contours and 
8744 Jy beam$^{-1}$ km s$^{-1}$ for the blue contours). 
Blue and red contours correspond  to blue- and red-shifted gas, 
with integrated velocity ranging from +20 to +58 km s$^{-1}$ 
and from +60 to +95 km s$^{-1}$,  respectively. 
The synthesized beam of the JVLA image is 0.28$''$ $\times$ 0.24$''$, 
whereas that of the SMA is 0.58$''$ $\times$ 0.43$''$.  
The dashed white box indicates the zoomed area plotted in the lower panel. 
{\it Lower Panel:} Overlay of the SO$_2$ [22$_{2,20}$ $\rightarrow$ 22$_{1,21}$] $0^{th}$ 
moment emission imaged with the SMA at 0\pas4 resolution (color image and black contours) 
with the (10,10) and (13,13) thermal \amm\,emission (blue and red contours, respectively),  
as well as several masers observed in W51-North.  
The integrated velocity range for SO$_2$ is from +50 to +70 km s$^{-1}$  
and the contours are from 30\% to 100\% with steps of 10\% of the line peak  
(7.8 Jy beam$^{-1}$ km s$^{-1}$). 
The \amm\,contours are 30\%, 60\%, 90\% of the line peak for the (10,10) line (66 mJy beam$^{-1}$ km s$^{-1}$)  
and the (13,13) line (65 mJy beam$^{-1}$ km s$^{-1}$). 
The black triangles mark the position of the water maser spots observed with the VLA,
while the white cross marks the centroid position of SiO masers imaged with the VLBA, 
($\alpha,\ \delta)_{J2000}$= (19$^h$ 23$^m$ 40\farcs055, 14$^\circ$ 31\arcmin ~5\farcs59), 
as reported in \citet{Eisner02}.  
The black cross marks the position of the \amm\,masers detected in this study with the JVLA. 
Note that the images displayed are not corrected for primary beam effects. 
}
\label{env}
\end{figure}

\begin{figure}
\includegraphics[angle=-90,width=0.5\textwidth]{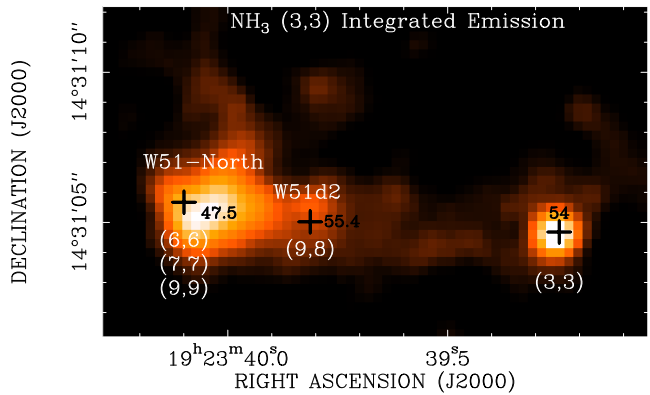}
\caption{ Overlay of the integrated flux of the \amm\,(J,K)=(3,3) line  (color image) 
with the positions of known \amm\,masers measured with interferometry (black crosses) in the W51~IRS2 region.
The \amm\,(3,3) map was made  with the VLA-C configuration at 1\pas2 resolution \citep{ZhangHo95}. 
The color scales are linear from 0 to 700 mJy beam$^{-1}$ km s$^{-1}$. 
Positions of the \amm\,masers from doublets (J,K)=(3,3) \citep{ZhangHo95}, (9,8) \citep{Wilson91}, (6,6), (7,7), (9,9) (this study), are indicated in the map. 
The numbers next to the crosses indicate  the maser peak velocities in \kms. 
}
\label{nh3_33}
\end{figure}

In order to characterize more in detail the environment of the \amm\,maser, 
and establish its relation with W51-North and the surrounding region, 
we can compare its position with respect to other 
dense gas and outflow tracers as well as different maser species.  
Figure~\ref{env} shows two overlays: 
the 25~GHz continuum emission (the same as shown in Figure~\ref{nh3+cont}) with contours of the SiO $v = 0$, $J = 5 - 4$  line emission from \citet{Zapata09} (upper panel) and 
 the image of the SO$_2$ $22_{2,20} - 22_{1,21}$ line from \citet{Zapata09} with contours of the (10,10) and (13,13) thermal \amm\,emission (lower panel). 
In these overlays, we also report positions of known maser species. 
Figure~\ref{nh3_33} shows an overlay of the integrated \amm\,emission from the (3,3) doublet with the positions of the known \amm\,masers in W51~IRS2 
 measured with interferometers. 

 The SiO 5-4 emission (imaged with the SMA at 0\pas4 resolution) shows the presence of a collimated bipolar outflow, 
 elongated northwest-southeast (PA$\sim150$\degree) over 4\arcsec\,or 22000 AU 
 and centered on W51-North (indicated by a  small plus in the upper panel of Fig.~\ref{env}). 
  The redshifted emission of the outflow (from +60 to +95 \kms) is located toward the northwest, while the blueshifted emission (from +20 to +58 \kms) points toward the southeast.
A similar spatial and velocity structure is probed by the 3D gas kinematics of 22 GHz water masers \citep{Imai02} and SiO masers \citep{Eisner02}, 
which trace the base of the SiO 5-4 bipolar outflow, at radii between 20 and 2000 AU. 
In particular,  proper motion measurements of the H$_2$O masers 
(which generally trace shocked gas in massive protostellar outflows - \citealt{Goddi05}) 
revealed the presence of a compact ($\sim$5000 AU) and fast ($\sim$100~\kms) outflow \citep{Imai02}. 
In addition, VLBI imaging of the SiO $v = 2 \ J = 1 - 0$ line indicates that SiO masers may be tracing the innermost 
and densest portions of this molecular outflow on scales as small as 50 AU \citep{Eisner02}. 
The latter interpreted the SiO distribution in terms of an accelerating outflow from a deeply-embedded high-mass YSO, 
with a 3D acceleration of 0.5 \kmsy\ (or 0.2 \kmsy\,projected along the line-of-sight). 

Based on the Effelsberg monitoring of several \amm\,maser lines, \citet{Henkel13} measured a velocity drift of the order of 0.2 \kmsy, for the 47 \kms\,NH$_3$ masers.  
Based on the remarkable agreement between both the velocity components and the velocity drifts of SiO and \amm\,masers, they proposed a physical relation between the two rare maser species. 
Our interferometric measurements  exclude that the two maser species are excited by the same YSO hosted by W51-North. 
In fact, 
in Figure~\ref{env} we have marked the positions of the 
water maser spots as well as the position of the SiO maser source reported by \citet{Eisner02} 
and we find that the \amm\,maser emission  is offset by 0\pas65 from the SiO masers and the centroid of the \wat\,masers.  
 We explicitly note that this offset is significantly larger than the positional accuracy or registration error between VLA measurements at 7~mm and 1~cm, $<<$0\pas1 (\citealt{Eisner02}; this work).  
Therefore, the \amm\,masers are not excited in the powerful molecular outflow probed at very small scales by SiO and H$_2$O masers and at large scales in thermal SiO (5-4) emission.  

What is then exciting these high-JK \amm\,masers?
Besides one single water maser spot (shown in the lower panel of Fig.~\ref{env}), there are no other known maser species spatially coincident with these \amm\,masers.  
In fact, the previously known \amm\,(3,3) masers  are excited West of W51d2 (offset by $\sim$9\arcsec; see Figure~\ref{nh3_33}), 
emit at V$_{\rm LSR}$ = 54~\kms, and show spatially extended emission \citep{ZhangHo95}. 
\citet{Wilson91} imaged the (9,8) \amm\,maser at V$_{\rm LSR}$ = 55.4~\kms\,with 0\pas2 resolution using the VLA and located its emission towards W51d2 (Figure~\ref{nh3_33}).  
\met\,masers, a typical signpost of high-mass star formation,  are  excited towards W51d2 but not in W51-North  \citep{Surcis12}. 
In spite of the lack of other maser species, there is strong thermal emission from high-density gas tracers at the position of the \amm\,masers.

The lower panel of Figure~\ref{env} shows the SO$_2$ $22_{2,20} - 22_{1,21}$ line  emission, imaged with the SMA at 0\pas4 resolution. 
Remarkably, the overlay reveals that the \amm\,maser corresponds to the strongest peak of SO$_2$.  
In Figure~\ref{env}, we also overlay the velocity-integrated (thermal) emission of the (10,10) and (13,13) \amm\,lines  
supposedly picking out the hottest thermal gas. 
Interestingly, this highly excited thermal emission shows two main condensations, 
the dominant one towards W51-North and the SiO/H$_2$O masers, 
and a second peak at the \amm\,maser position. 
It is worth noting that the peak emission  of the 47-50~\kms\,\amm\,masers is also the peak
         position of other high excitation lines, 
like HCOCH$_2$OH (62$_{13,49}$-62$_{12,50}$) with $E_l=1476$~K and HC$_3$N (38-37) ($\nu_t$=0) with $E_l=306$~K \citep{Zapata10}. 
 \citet{Zapata09} proposed that the SO$_2$ emission traces a large and flattened molecular ring, 
 (with an inner cavity of about 3000 AU and size of about 9000 AU), seen nearly face-on,  
 surrounding the dusty disk\footnote{This cavity is observed in other submm lines with moderate excitation above the ground state, 
 like SO ($8_9-7_8$), SO$_2$ $19_{1,19} - 18_{0,18}$, HC$_3$N (38-37), and C$_2$H$_5$OH ($37_{8,29}-36_{9,28}$).}. 
 
\citet{Zapata10} speculated on the possibility that W51-North may be a binary system. 
The potential warm ``companion'' of W51-North would be located at the SO$_2$ peak to the northeast of the disk/ring. 
Without a clear evidence of outflowing gas activity associated with this potential ``companion'' (from SiO 5-4 or CO 3-2), 
 the molecular peak could still be in principle the result of the interaction of the outflow from W51-North with a high density portion of the molecular clump, as opposed to be a distinct YSO.  
\citet{Zapata10} however noticed that the blueshifted lobe of the CO ($J=3-2$) outflow has a different position angle with respect to the redshifted
lobe (see their Figure 1), indicating possibly the presence of multiple outflows or at least of a second outflow. 
In fact, the presence of a binary system (although more compact) at the center of W51-North has been also suggested by the precession of the outflow at very small scales as traced by \wat\,and SiO masers \citep{Eisner02}. 
Our measurements indicate that the size of the cavity observed in SO$_2$ corresponds to the distance between the SiO and \amm\,masers.  
If the molecular ring traces a circumbinary disk, seen face-on, then the two maser species would pinpoint the positions of two high-mass YSOs 
accreting and rotating in the same circumbinary disk:  
the SiO masers would pinpoint the main member, W51-North, which is driving the powerful outflow, and the \amm\,masers would locate the companion.
The latter could be a consequence of disk fragmentation due to gravitational instabilities expected in a large, massive disk \citep[e.g.,][]{kratterMatzner06}. 
We conclude that, although  the two rare \amm\,and SiO masers are not excited in the same YSO, they may have a close physical association in a binary.  
This physical connection may also explain the fact that  both maser species show a velocity component around 47~\kms\,and a secular velocity drift of  $\sim$0.2~\kmsy.  
 We warn that the binary scenario is plausible  based  on the morphology of dense gas and the positions/velocities of individual maser species, but is inconsistent with the velocity field of the thermal emission (neither the SO$_2$ nor the \amm\,lines show a component at 47~\kms). 

Is there any direct indication of an outflow from the companion?
The double-peaked profile coupled with the North-South elongation of the \amm\,(7,7)
         maser region could in principle trace a small protostellar disk or a slow outflow. 
In a circumstellar disk observed nearly edge-on, the gain paths of the redshifted receding and the blueshifted approaching components of the disk are sufficiently long  to significantly amplify the background emission at the corresponding velocities. 
In fact, double-peaked maser profiles based on single-dish spectra have been sometimes interpreted  in the framework of rotation in Keplerian disks in the past \citep[e.g. see][in the case of S255-IR]{Cesaroni90}. 
In specific cases, however, VLBI imaging revealed that these double-peaked profiles were not due to rotation but outflow \citep[e.g. see][again in the case of S255-IR]{Goddi07}. 
In the case of W51-North a disk is more improbable, because the velocity distribution of individual maser spots is not symmetric with respect to  the LSR velocity of the source, i.e. 47.5~\kms\,(assuming this corresponds to the central velocity of the 6,6 or 9,9 lines).  
We prefer the outflow interpretation, based on two additional arguments. 
First, the (7,7) emission is elongated perpendicular to the East-West (circumbinary) disk observed in SO$_2$. 
Second, the more redshifted features are located in the North (and the more blueshifted in the South), 
which is consistent with the geometry and velocity of the SiO 5-4 outflow driven by W51-North. 
If the \amm\,masers are truly tracing a binary member accreting in the same circumbinary disk along with W51-North, 
then  in principle the two outflows could share a similar orientation and the large-scale outflow (traced by thermal SiO and CO emission) 
could receive contributions from both YSOs. 
Nevertheless, to discriminate between the different possibilities (disk, outflow, or even multiple distinct gas components), 
higher spatial resolution observations of the \amm\,(7, 7) line with VLBI will be required to resolve and image the ammonia maser emission. 
The fact that we do not see such a double-peaked profile in the \amm\,(6, 6) or (9,9) lines may be explained by different physical/kinematic structures traced by ortho and para-\amm\,(the two behave like different molecular species - \citealt{Cheung69}).  

It remains to understand the velocity offset between the maser ($\sim$47~\kms) and thermal emission (centered at 60~\kms). 
A similar behavior has been seen in NGC6334I, where the (6,6) maser line is offset by 13 ~\kms\,
from the thermal emission at the same position \citep{Beuther07}. 
Based on the elongated morphology, \citet{Beuther07} proposed that the (6,6) \amm\,maser may be associated with some outflow-shock processes, 
similar to what we propose here for the (7,7) \amm\,masers in W51-North.

\subsection{Constraints on the pumping mechanism of the \amm\,masers}

What is exciting these \amm\,maser lines? 
The pumping mechanisms for most ammonia masers are still unclear. 
\citet{Henkel13} describes three main pumping schemes  for creating a population inversion of the ammonia upper levels: 
(1) collisions; (2)   overlap with specific molecular lines; and (3)    infrared continuum radiation from dust. 

Metastable transitions of ortho-ammonia, like (3,3), are known to be collisionally excited \citep{Walmsley83}. 
Since the $K =0$ rotational states are not split into inversion levels due to nuclear spin statistics, 
and collisional transitions involving parity changes between $K=0$ and $K=3$ levels are more favorable, 
the upper level of the (3,3) doublet connected with the (0,0) state can be overpopulated with respect to the lower level of the (3,3) doublet.  
Although more complicated, similar arguments could justify a maser in higher excitation lines like the (6,6) or the (9,9).  
In this scheme, however, masers in para-ammonia lines such as the (7,7) line would be hard to explain. 
Another complication is that collision rates are only known for inversion levels $J \leq 6 $ \citep[e.g.,][]{Danby88}, 
and extrapolation to higher J is a large source of uncertainty. 
Finally,  most of the masers observed by \citet{Henkel13} are non-metastable,  
and extremely high densities ($10^{10}-10^{12}$ \cmc)  would be required if they were collisionally excited. 
 Therefore the extreme density requirements argue against their collisional excitation.
If the non-metastable masers detected by \citet{Henkel13} were physically associated with the metastable masers imaged here, 
we consider it unlikely that collisional excitation is at work here.

A fortuitous overlap of a far-infrared line, which allows a population transfer, for example, from a (J, K) inversion level to (J + 1, K), is another possibility.
 Since from the single-dish study, there appear to be 19 distinct maser transitions, 
 it is unlikely that {\it all} of these lines are due to several chance alignments of molecular transitions
 allowing inversion population in these metastable {\it and} non-metastable states.
In addition, based on morphological arguments, it may be the case that there is a difference between masers in ortho and para-\amm\,(see Sect.~\ref{spots_maps}),  
indicating that the \amm\,maser excitation may not result from an overlap with lines from other molecules.  
Therefore, an alternative mechanism to collisions and line overlap is required.  

We believe that the pumping mechanism for ammonia masers in W51-North must be via infrared photons. 
\citet{Madden86} was the first to suggest a strong infrared radiation field, such as found around a deeply embedded high-mass YSO, as a pumping         mechanism for non-metastable transitions. 
The kinetic temperature of the bulk of the ammonia emitting gas in W51~IRS2 is $\sim$300 K, as derived by single-dish observations \citep{Mauersberger87},  
which is comparable to the kinetic temperature of the molecular ring, 250~K,  measured with interferometers by \citet{Zapata10} in W51-North. 
This high temperature (at densities where dust and gas temperatures are likely coupled) facilitates vibrational excitation by infrared photons near 10~$\mu$m, which correspond to the intensity peak of a black body at \tkin$\sim$300~K.  
Therefore, we believe there is a relation of the observed masers to the ammonia vibrational levels. 
Vibrational excitation may cause significant deviations from quasi-thermal conditions in the ground state.
Based on the rapid time variability \citep{Henkel13}, the masers must have an exponential gain, 
e.g. they may be  amplified by a continuum background. 
In this respect, W51-North, showing a combination of hot dust and molecular gas, may be the ideal location for ammonia masers. 

\citet{Schilke89} built an LVG code, where he extrapolated the \amm\,collision rates to high-J ($>$6) levels and included also vibrational excitation by radiation (the $v2 = 1$ level).  
The code  predicts the formation of a large number of maser lines at  \tkin$>$240K,  a temperature at which a sufficient number of  10~$\mu$m  photons is indeed produced.  
 The problem however is that predictions from current models are highly speculative. 
 For example, \citet{BrownCragg91} built a model that did not have gas and dust well
mixed, but had only external dust adjacent to gas, which is clearly not realistic. 
In addition, the silicate dust feature around 10~$\mu$m may considerably  alter the radiation field. 
As modeling work in progress, we plan to modify the IR spectrum testing different shapes of the silicate feature, which may affect the vibrational ground state ammonia spectrum. 



\section{Summary}

We have mapped the \amm\,emission from the (6,6), (7,7), (9,9), (10,10), and (13,13) inversion transitions in W51-North with 0\pas2 resolution.  
The main findings of this paper are listed below:

\begin{itemize}

\item
 We detected both extended thermal ammonia emission in five inversion lines, with rotational states ranging in energy from about 400 to 1700~K, 
 and point-like ammonia maser emission in the (6,6), (7,7), and (9,9) transitions.  
\vspace{0.2 cm} 
 \item 
 We established a spatial and velocity correlation of three metastable maser transitions of ammonia, 
demonstrating that, although the two ortho and the one para lines are associated with the same object, 
they arise from slightly different volumes of gas. 
\vspace{0.2 cm} 
 \item
  The \amm\,(7, 7) emission is not single-peaked but it shows a double-horn profile, and it is extended across 40 AU along north-south. With the current data we cannot differentiate whether this double-horn line profile 
  may be the signature of a potentially underlying accretion disk or associated with a bipolar outflow, although the latter hypothesis is more likely 
  based on the velocity distribution of maser spots. 
  Higher spatial resolution observations of the \amm\,(7, 7) line with VLBI are required to address this question.   
\vspace{0.2 cm} 
\item
 We established that the \amm\,maser is separated (by 0\pas65) from the emission centroids of H$_2$O  and SiO masers, the latter
 tracing a high velocity bipolar outflow emanating from W51-North. 
 This excludes that the \amm\,masers are directly excited by W51-North. 
\vspace{0.2 cm} 
\item
Our \amm\,maser maps favor the presence of a multiple protostellar system, or at least of a binary, in W51-North:  
the dominant YSO associated with SiO and \wat\,masers and a  companion pinpointed by the \amm\,maser. 
This finding, along with the fact that  both maser species show a velocity component around 47~\kms\,and a secular velocity drift of  $\sim$0.2~\kmsy, 
suggests that the two rare \amm\,and SiO masers may have a close physical association in a binary.  
\vspace{0.2 cm} 
\item
We believe that the pumping mechanism for ammonia masers in W51-North must be via infrared photons.

\end{itemize}

\noindent
Establishing the location of the ammonia masers in W51~IRS2 has provided interesting insights on  \amm\,maser excitation
and our general understanding of W51-North, which gives rise to rare \amm\,and SiO masers. 
The detection of many \amm\,maser lines, from both metastable and  non-metastable transitions of ortho and para ammonia,
makes W51-North  a unique target for ammonia maser research. 
Interferometric studies of all the known \amm\,masers in W51, similar to the one presented here, 
will provide important constraints for the development of theoretical models on ammonia maser pumping. 

\begin{acknowledgements} 
 We thank Dr. J. Eisner for providing the positions and velocities of water maser spots detected in W51-North with the VLA. 
We thank the anonymous referee for a constructive report. 
We are grateful to Malcolm Walmsley for carefully reading the manuscript and for providing useful suggestions. 
\end{acknowledgements}

\bibliographystyle{aa}
\bibliography{biblio}  

\end{document}